\documentclass[12pt]{article}
 \normalsize
\begin{document}
\begin{center}
{\bf{NON-LINEAR SIGMA MODEL ON CONIFOLDS}}

\vspace{1.0cm}

R.Parthasarathy{\footnote{Permanent address: The Institute of Mathematical 
Sciences, Chennai 600113, India: e-mail address:pragavac@sfu.ca and 
sarathy@imsc.ernet.in}} and 
K.S.Viswanathan{\footnote{e-mail address: kviswana@sfu.ca}} \\
Department of Physics, Simon Fraser University \\
Burnaby.B.C., Canada V5A 1S6. \\

\end{center}

\vspace{0.5cm}

{\noindent{\it{Abstract}}

\vspace{0.5cm}

Explicit solutions to the conifold equations with complex dimension $n=3,4$ 
in terms of {\it{complex coordinates (fields)}} are employed to construct 
the Ricci-flat K\"{a}hler metrics on these manifolds. The K\"{a}hler 
2-forms are found to be closed. The complex realization of these conifold 
metrics are used in the construction of 2-dimensional non-linear sigma model 
with the conifolds as target spaces. The action for the sigma model is shown 
to be bounded from below. By a suitable choice of the 'integration constants', 
arising in the solution of Ricci flatness requirement, the metric and the 
equations of motion are found to be {\it{non-singular}}. 
As the target space is Ricci-flat,
the perturbative 1-loop counter terms being absent, the model becomes 
topological. The inherent $U(1)$ fibre over the base of the conifolds is shown 
to correspond to a gauge connection in the sigma model.  

The same procedure is employed to construct the metric for the resolved 
conifold, in terms of complex coordinates and the action for a non-linear 
sigma model with resolved conifold as target space, is found to have a 
minimum value, which is topological. The metric is expressed in terms of the 
six real coordinates and compared with earlier works. The harmonic function,
which is the warp factor in Type II-B string theory, is obtained and the 
ten-dimensional warped metric has the $AdS_5 \times X_5$ geometry. 

\newpage 	
{\noindent{\bf{1. Introduction}}}

\vspace{0.5cm}

The original motivation behind the construction of conifolds was to find a non-trivial Calabi-Yau manifold to be used as the transverse space to Minkowski space in superstring theory, instead of the usual flat transverse space. In 1997, Maldacena [1] showed that the large $N$ limit of ${\cal{N}}=4\ SU(N)$ gauge theory
is related to Type II-B string theory on $AdS_5\times S^5$. ($S^5$ 
preserves maximal number of supersymmetries, namely 32) Subsequently, other
backgrounds such as $AdS_5\times X^5$ for Type II-B theory have been explored 
with $X^5$ as Einsteinian, and are found to be related ${\cal{N}}=1$ superconformal theory in four dimensions [2]. A new example of duality was found by Klebanov and Witten [3] where the $X^5$ is a smooth Einstein manifold and a coset 
space such as the one considered in the context of Kaluza-Klein theory [4,5].
The
remarkable role played by conifolds acquired recent interest due to the 
observation of Klebanov and Strassler [6] that certain warped product, with
the warp factor depending on the radial coordinate of the conifold, have the 
interpretation of D3-brane solution of supergravity. This is further confirmed
by Herzog, Klebanov and Ouyang [7]. D3-branes on conifold points have been
studied [8,9] and related to that of a Type II-B supergravity on a 
non-compact Calabi-Yau 3-fold which is a deformed conifold, topologically
a 6-dimensional cone over $S^2\times S^3$ base, apex being replaced by $S^3$.
Wrapping of branes on 2-sphere of a conifold leading to fractional D-branes
has been considered by Papadopoulos and Tseytlin [10]. A conformal field theory  from Calabi-Yau 4-folds is obtained by Gukov, Vafa and Witten [11]. Further,
a strong structural similarity between the deformed conifold and 
the moduli space of $CP^1$-lumps
of unit charge is recently poited out by Speight [12]. Using the Wilson loop 
construction of Maldacena [1], to an $AdS_5\times X^5$ background, Caceres
and Hernandez [13] study the Higgs phase of large N field theory on the conifold and are able to relate quark-antiquark interaction to the parameter in the
harmonic function (warp factor) in the near horizon limit. 

\vspace{0.5cm}

While these studies provide a spectacular progress in our understanding of
the Calabi-Yau compactification of string theory, it is worth recalling the
observation of Strominger [14] that the low energy effective field theories
arising from Calabi-Yau compactification are generally inconsistent or 
ill-defined at the classical level due to the conifold singularities, which
arise in any Calabi-Yau compactification. Low energy effective theory, as
described by sigma-model whose target space is the moduli space (or equally the
conifold, see [12] for their structural similarity), has the equations of motion becoming singular. Therefore, it is important to first understand non-linear
sigma models whose target space is a conifold and this is one of the purposes of 
this study. 

\vspace{0.5cm}

An explicit construction of the metric on Ricci-flat K\"{a}hler manifold, was
first (to the best of our understanding) made by Candelas and de la Ossa [15]
in terms of real coordinates. Non-singular Ricci flat K\"{a}hler metrics which
generalize the Eguchi-Hanson metric were constructed by Stenzel [16] following
the method of Gibbons and Pope [17]. Harmonic forms and Brane resolutions  
on these manifolds are recently constructed by Cvetic, Gibbons, Lu and
 Pope [18]. 
One observation in [18] namely, it is possible to choose
the integration constants such that the metric is non-singular and the horizon
can be completely eliminated, will be borne out by our explicit construction.
This in turn allows us to realize non-linear sigma model on conifolds with the
equations of motion not becoming singular at small distances, thereby avoiding
the difficulty alluded by Strominger [14]. The construction of the Ricci-flat 
K\"{a}hler metric is based on the strategy: An $O(n)$ invariant quantity  
is first constructed and the K\"{a}hler potential is assumed to be a differentiable function of
this invariant. The components of the metric are evaluated in terms of the 
first and second order differentials of this function with respect to the 
invariant. The determinant of the resulting metric is equated to a constant 
times $|{\cal{F}}|^2$, where ${\cal{F}}$ is a holomorphic function.  
This results in a non-linear differential equation for
the function. It is enough to solve this for the first order differential of 
the function, thereby determining the Ricci-flat metric. 
This procedure introduces integration constants.
The K\"{a}hler 
structure is then verified. This strategy is followed in [19] to construct the
metrics on conifold. An alternative procedure is by working out covariantly
constant spinor equations [20] as this implies Ricci flatness. In this paper,
we follow the differential equation method. In our study, we employ complex
coordinates for the conifold, which is not the case in the earlier works [15,
19]. 
This is partly due to our earlier usage of these
coordinates to describe quadrics in $CP$-space [21] guided by the 
construction of generalized Gauss maps to describe 2-dimensional surfaces
conformally immersed in $R^n$ [22]. The use of complex coordinates facilitates
the setting up of the action for non-linear sigma model with the target space
as conifold, which is our second motivation. The action for this sigma model   
will be shown to have a lower bound and the equations of motion do not 
become singular by a choice of the 'integration constants'.  

\vspace{0.5cm}

After the Ricci-flat K\"{a}hler metric construction  
in n=2 and 3 complex dimensional conifolds, and their role in non-linear sigma model, we consider the case of resolved conifold, using the complex coordinates, as 
our third motivation. 
We rewrite the metrics in
terms of real coordinates to compare with the earlier results. This gives a better
understanding of the conifolds. In particular we clearly see how the choice of integration
constants avoid the singularity in conifold in comparison with the resolved conifold. The
additional sphere introduced to replace the apex of the conifold in the resolved conifold
case, corresponds to the non-zero value of one of the integration constants in
the ordinary conifold case.  

\vspace{0.5cm}

{\noindent{\bf{2. Ricci-Flat K\"{a}hler Metric on Conifold}}}

\vspace{0.5cm}

An $O(n)$ symmetric target space with n-complex fields (which can be thought of as
coordinates) satisfying the equation,
\begin{eqnarray}
\sum_{i=1}^{n} ({\phi}^i)^2 &=& 0,
\end{eqnarray}
is a conifold of real dimension $(2n-2)$. It is a smooth manifold except for the point
${\phi}^i=0$. If ${\phi}^i$ solves (1), so also $\psi {\phi}^i$ for any $\psi\ \in\
C$ and so, the surface (1) is made up of complex lines passing through an origin,
thereby giving the conical structure. This surface admits the U(1) symmetry group
, namely, ${\phi}^i\rightarrow e^{i\theta}{\phi}^i$. The point ${\phi}^i=0$ is a 
double point singularity. The points ${\phi}^i$ and 
$\psi\ {\phi}^i$ are not to be identified. The base of the conifold is given
by the intersection of the space of solutions of (1) with a sphere in $C^n$,
\begin{eqnarray}
\sum_{i=1}^{n}|{\phi}^i|^2 &=& r^2.
\end{eqnarray}
If the right side of (1) is replaced by $a^2$, then we get a 'deformed conifold' and the metric
on this has been obtained in [15] in terms of real coordinates for the case n=4. The general
procedure followed is to assume that the metric can be obtained from a scalar differentiable
function of $r^2$, say $K(r^2)$. Then using complex coordinates, $g_{i\bar{j}}\ =\
 {\partial}_i{\partial}_{\bar{j}}K(r^2)$. The determinant of the metric is 
equated to a constant times square of the modulus of a field (coordinate)  
 so that the Ricci tensor, $R_{i\bar{j}}\ =\ 
-{\partial}_i{\partial}_{\bar{j}}\ell og\ det\ g$ vanishes. This procedure yields a second order differential equation for $K(r^2)$ and it will be enough to solve for $K'(r^2)\ =\ 
\frac{dK(r^2)}{dr^2}$ for finding out the metric components.  

\vspace{0.5cm}

For the (complex) $n$-conifold equation (1), we parameterize the $n$-complex 
fields ${\phi}^i$ satisfying (1), by,
\begin{eqnarray}
\{ {\phi}^1, {\phi}^2, \cdots {\phi}^n\} &=&\psi \bigl\{ 1-\sum_{i=1}^{n-2}
{\zeta}^2_i, i(1+\sum_{i=1}^{n-2}{\zeta}_i^2),2{\zeta}_1,2{\zeta}_2,
 \cdots 2{\zeta}_{n-2} \Bigr\},
\end{eqnarray}
in terms of $(n-1)$ complex fields $\psi, {\zeta}_1, {\zeta}_2,\cdots 
{\zeta}_{n-2}$. This parameterization has been suggested by Hoffman and 
Osserman [22] in their study of Gauss map of immersed surfaces in $R^n$ to 
describe the complex quadric and investigated by us [21] in our study 
of string theory in which the string world-sheet has been viewed as a 
2-dimensional surface in $R^n$ and also in curved space. The base of the 
conifold is given by the intersection of the space of solutions of (1), 
that is (3), with a sphere in $C^n$, given by (2). Using (3), we find,
\begin{eqnarray}
r^2 &=& 2|\psi |^2 \Bigl\{ 1+2\sum_{i=1}^{n-2}|{\zeta}_i|^2 + 
| \sum_{i=1}^{n-2}{\zeta}^2_i|^2 \Bigr\}.
\end{eqnarray}

\vspace{0.5cm}

In our parameterization, the limit $r \rightarrow 0$ 
corresponds to $\psi \rightarrow 0$. The 'strategy' of finding the Ricci 
flat metric on the $n$-conifold yields,
\begin{eqnarray}
g_{\psi\bar{\psi}}&=&\frac{a(n-2)}{(n-1)|\psi |^2} (ar^{2n-4}+b)^{\frac
{2-n}{n-1}}\ r^{2n-4}, \nonumber \\
g_{\psi {\bar{\zeta}}_i}&=&\frac{a(n-2)}{n-1}\ {\bar{\psi}} (ar^{2n-4} 
+b)^{\frac{2-n}{n-1}}\ r^{2n-6}\ \frac{\partial Y}{\partial {\bar{\zeta}}
_i}, \nonumber \\
g_{{\zeta}_i{\bar{\zeta}}_j}&=&\frac{a(n-2)}{(n-1)|\psi|^4}\ (ar^{2n-4} 
+b)^{\frac{2-n}{n-1}}\ r^{2n-8} \frac{\partial Y}{\partial {\zeta}_i} 
\ \frac{\partial Y}{\partial {\bar{\zeta}}_j} \nonumber \\
&+& (ar^{2n-4}+b)^{\frac{1}{n-1}}\Bigl\{ \frac{1}{Y}\frac{{\partial}^2Y}
{\partial {\zeta}_i \partial {\bar{\zeta}}_j} - \frac{1}{Y^2} \frac{
\partial Y}{\partial {\zeta}_i}\ \frac{\partial Y}{\partial {\bar{\zeta}}
_j} \Bigr\},
\end{eqnarray}
where $Y\ =\ r^2/{|\psi|^2}$, $a$ and $b$ are integration constants. The 
metric becomes singular when $\psi = 0$, even when $b \neq 0$. This is removed 
by redefining $\psi$ as $\Psi \ =\ \frac{{\psi}^{n-2}}{n-2}$. Then as long 
as $b \neq 0$, the metric is smooth. A non-linear sigma model on the target 
space with the metric (4) can be constructed by taking the $n-1$ complex 
fields as functions of $z$ and $\bar{z}$, the 2-dimensional space, by the 
action,
\begin{eqnarray}
S&=& \int g_{\alpha {\bar{\beta}}} \Bigl\{ {\partial}_z \alpha {\partial} 
_{\bar{z}}{\bar{\beta}}+{\partial}_{\bar{z}}\alpha {\partial}_z\bar{\beta}
\Bigr\} \Big( \frac{i}{2}\Big) dz d\bar{z},
\end{eqnarray}
where $\alpha$ and $\beta$ stand for $\psi, {\zeta}_1, \cdots {\zeta}_{n-2}$. 
The K\"{a}hler two form $\omega \ =\ -2i g_{\alpha \bar{\beta}} d\alpha 
\wedge d\bar{\beta}$ is found to be closed and so the integral of this 
will be a topological invariant. This is used to show that the action (5) is 
bounded from below and the minimum action is topological. We will now 
consider explicitly the cases for $n=3$ and $n=4$ and show the above results. 

\vspace{0.5cm}

{\noindent{\it{2.1: $n=3$ Conifold}}

\vspace{0.5cm}

The $O(3)$ symmetric target space with three complex fields satisfying  
\begin{eqnarray}
\sum_{i=1}^{3} ({\phi}^i)^2 &=& 0,
\end{eqnarray}
is a $3$-conifold. The above equation is solved by the parameterization
equations, 
\begin{eqnarray}
{\phi}^1\ =\ \psi(1-f^2); & {\phi}^2\ =\ i\psi(1+f^2);& {\phi}^3\ =\ 2\psi f,
\end{eqnarray}
where $\psi$ and $f$ are complex coordinates.  
In our description of two dimensional  
non-linear sigma model, these
will be taken as functions of $z$ and $\bar{z}$.  
The invariant $r^2$ becomes,
\begin{eqnarray}
r^2 &=& 2{|\psi|}^2(1+{|f|}^2)^2.
\end{eqnarray}
Now using (7) and 
(8), and following the 'strategy' outlined in the Introduction, the metric on the 
conifold is obtained in terms of the complex fields $\psi, f$ basis as
\begin{eqnarray}
g&=& \left( \begin{array}{lcr}  
\frac{Ar^2}{2{|\psi |}^2\sqrt{Ar^2+b}} &\ & \frac{2A\bar{\psi}f(1+{|f|}^2)}
{\sqrt{Ar^2+b}} \\
  &\ & \\
\frac{2A\psi \bar{f}(1+{|f|}^2)}{\sqrt{Ar^2+b}} &\ & \frac{4A{|\psi |}^2
(1+{|f|}^2)}{\sqrt{Ar^2+b}}+\frac{4{| \psi |}^2b}{r^2\sqrt{Ar^2+b}} \\
 \end{array} \right) 
\end{eqnarray}    
Here $A$ and $b$ are integration constants ($A\ \neq 0$, $b$ could be zero). The determinant
of the metric is $2A$, thereby making the manifold Ricci flat. The metric is hermitian and 
satisfies,
\begin{eqnarray}
\frac{\partial g_{\alpha \bar{\beta}}}{\partial \gamma}\ =\ \frac{\partial g_{\gamma
\bar{\beta}}}{\partial \alpha} &;& \frac{\partial g_{\alpha \bar{\beta}}}{\partial
\bar{\gamma}}\ =\ \frac{\partial g_{\alpha \bar{\gamma}}}{\partial \bar{\beta}},
\end{eqnarray}
where $\alpha, \beta$ stand for $\psi, f$. Conditions (11) are 
necessary and sufficient [23] to show that  
the K\"{a}hler 2-form
$\omega $,
\begin{eqnarray}
\omega &=& -2i\sum_{\alpha, \beta} g_{\alpha \bar{\beta}}\ d\alpha \wedge d\bar{\beta},
\end{eqnarray}
is {\it{closed}}. This property will be used to obtain a lower bound for the action of
the non-linear sigma model with target space described by the above metric. 

\vspace{0.5cm}

When the integration constant $b\neq 0$, in the limit $r\rightarrow 0$, 
i.e., $\psi \rightarrow 0$ 
, the above metric (10) for the conifold, does not become singular. 
This explicit
behaviour is mentioned in the Introduction and is pointed out in [18].   

\vspace{0.5cm}

To appreciate the role played by the integration constant $b$ in not making
the metric singular at short radial distances, we give the metric in terms 
of real coordinates. We choose the following representation: $\psi \ =\ 
\frac{r}{\sqrt{2}}e^{2i\phi}\ cos^2(\frac{\theta}{2})$ and $f\ =\ e^{i\xi}
tan(\frac{\theta}{2})$ such that $2|\psi|^2(1+|f|^2)^2 = r^2$. Then working out the metric, we find that the above metric (10) corresponds to
\begin{eqnarray}
(ds_4)^2&=&\frac{A}{2\sqrt{Ar^2+b}} (dr)^2+\frac{r^2}{2}\{ (d\theta)^2 + 
sin^2{\theta} (d\xi )^2\} \nonumber \\
&+&\frac{r^2}{4}{\Bigl\{d\tilde{\phi} + cos{\theta}d\xi\Bigr\} }^2
+\frac{b}{2\sqrt{Ar^2+b}} 
\{ (d\theta)^2 + sin^2\theta (d\xi)^2\},
\end{eqnarray}
where we have introduced $\tilde{\phi}=-2\phi-\xi$. 
Such a geometrical construction is called 'resolving the 
singularity' 
(which will be studied in Section.3).  
The integration constant $b$ can be 
interpretted as 'resolution parameter'.  

\vspace{0.5cm}

We now consider bosonic non-linear sigma model in two dimensions with the 
target space as the conifold with metric (10). 
The action for this model is
given by,
\begin{eqnarray}
S&=& \int \Bigl\{ g_{\psi \bar{\psi}}({\psi}_z{\bar{\psi}}_{\bar{z}}+
{\psi}_{\bar{z}}{\bar{\psi}}_z)+g_{\psi\bar{f}}({\psi}_z{\bar{f}}_{\bar{z}}
+{\psi}_{\bar{z}}{\bar{f}}_z) \nonumber \\
&+&g_{f\bar{\psi}}(f_{\bar{z}}{\bar{\psi}}_z+f_z{\bar{\psi}}_{\bar{z}})+
g_{f\bar{f}}(f_z{\bar{f}}_{\bar{z}}+f_{\bar{z}}{\bar{f}}_z)\Bigr\}\ \Big(
\frac{i}{2}\Big) dz\ d\bar{z},
\end{eqnarray}
where the metric components $g_{\psi \bar{\psi}}, g_{\psi \bar{f}}, 
g_{f\bar{\psi}}, g_{f\bar{f}}$ can be read-off from (9), and the subscripts
$z,\bar{z}$ denote partial derivatives with respect to $z$ and $\bar{z}$ 
respectively.

\vspace{0.5cm}

The integral of the K\"{a}hler 2-form $\omega$ is
\begin{eqnarray}
\int \omega &=& c\int \Bigl \{ g_{\psi \bar{\psi}}({\psi}_z{\bar{\psi}}
_{\bar{z}}-{\psi}_{\bar{z}}{\bar{\psi}}_z)
+g_{\psi \bar{f}}({\psi}_z{\bar{f}}_
{\bar{z}}-{\psi}_{\bar{z}}{\bar{f}}_z) \nonumber \\
&+& g_{f\bar{\psi}}(f_z{\bar{\psi}}_{\bar{z}}-f_{\bar{z}}{\bar{\psi}}_z)
+g_{f\bar{f}}(f_z{\bar{f}}_{\bar{z}}-f_{\bar{z}}{\bar{f}}_z)\Bigr\}\ 
 \Big(\frac{i}{2}\Big) dz d\bar{z},
\end{eqnarray}
which, in general, is not a topological invariant. However, in view of the fact 
that the above 2-form $\omega$ in (12) is closed, $\int \omega$ is 
indeed a topological invariant [24]. Then it follows from (14) and (15) that,
\begin{eqnarray}
S+c^{-1}\int \omega &=& 2\int \Bigl\{ g_{\psi \bar{\psi}}{|{\psi}_z|}^2
+g_{\psi \bar{f}}{\psi}_z {\bar{f}}_{\bar{z}} \nonumber \\
&+& g_{f\bar{\psi}} f_z {\bar{\psi}}_{\bar{z}}+g_{f\bar{f}}{|f_z|}^2
\Bigr\} \big( \frac{i}{2}\big) dz d\bar{z}.
\end{eqnarray}
The integrand on the right hand side of (16) can be simplified using (10),
and after some algebra, we find,
\begin{eqnarray}
S+c^{-1}\int \omega &=& 2\int \Bigl\{ \frac{A(1+|f|^2)^2}{\sqrt{Ar^2+b}}
\left |\ {\psi}_z + \frac{2\psi \bar{f} f_z}{1+|f|^2}\ \right |^2  \nonumber \\
&+& \frac{4|\psi|^2}{r^2}\sqrt{Ar^2+b} |f_z|^2 \Bigr\} \frac{i}{2} dz d\bar{z},
\nonumber \\
&\geq & 0.
\end{eqnarray}
Repeating the steps for $S-c^{-1}\int \omega$, we have,
\begin{eqnarray}
S &\geq & |c^{-1}\int \omega|,
\end{eqnarray}
which guarantees  minimum for the action $S$. The equality holds good when
the fields $\psi, f$ are either holomorphic or 
anti-holomorphic, in which case the minimum
of the action $S$ corresponds to a topological invariant. In view of 
the observation that the metric does not become singular as $\psi 
\rightarrow 0$ (at the conical singularity)  as long as $b\neq 0$, the 
classical equations of motion for the action $S$ do not become singular,
thereby circumventing the criticism of Strominger [14].  

In contrast
to the $CP^n$-models, this invariant need not be an integer. In the limit
$\psi \rightarrow 0$, the metric
is not singular and the above action (14) reduces to,  
\begin{eqnarray}
S_{\psi \rightarrow 0}&=& 2\sqrt{b}\int \frac{f_z{\bar{f}}_{\bar{z}}+f_
{\bar{z}}{\bar{f}}_z}{(1+|f|^2)^2},
\end{eqnarray}
the standard action for $CP^1$ non-linear sigma model. The target space of 
this sigma-model is the base of the conifold, which can be thought of as a
quadric of dimension $2\sqrt{b}$. Interestingly, this action coincides with 
 the extrinsic curvature action for the string world-sheet,  
 without the integrability conditions, studied in [21]. The 1-loop 
partition function of this model [25] corresponds to that of a 2-dimensional
Coulomb gas. 

\vspace{0.5cm}

Now, returning to the minimum action in (18), which can be written in a 
compact notation as,
\begin{eqnarray}
S&=& \int g_{\alpha \bar{\beta}} {\partial}_z{\phi}^{\alpha}\ {\partial}
_{\bar{z}}{\bar{\phi}}^{\beta}\ \frac{i}{2} dz d\bar{z},
\end{eqnarray}
the perturbative quantum 1-loop effects can be studied by fluctuating the 
fields $\psi$ and $f$ from their classical values. As the metric $g_{\alpha 
\bar{\beta}}$ involves $\psi$ and $f$, exercising care for the covariant 
structure, it has been shown [26] that the 1-loop counter term must be of 
the form $T_{\alpha \bar{\beta}}\ =\ a_1 R_{\alpha \bar{\beta}} + a_2 R 
g_{\alpha \bar{\beta}}$. As the target space on which the sigma model is
defined, is {\it{Ricci flat}}, it follows that there will be no perturbative 
1-loop corrections. Thus the non-linear sigma model on the conifold (7) 
is a topological field theory and is scale invariant.  

\vspace{0.5cm}

From (17), we see that the term 
${\psi}_z+\frac{2\bar{f}f_z}{(1+|f|^2)}\psi$ can be
rewritten as a covariant derivative $D_z\psi \ \equiv \ ({\partial}_z+ {\cal{A
}}_z)\psi$, with the gauge connection ${\cal{A}}_z\ =\ \frac{2\bar{f}f_z}{
(1+|f|^2)}$. The appearance of gauge connection within the theory is not 
surprising, as this corresponds to the $U(1)$ fibre over $S^2$ as exemplified
 in (13). So, we are able to see the role of this $U(1)$ fibre over $S^2$ in the
action for the non-linear sigma model. For the case of the fields being
holomorphic, this gauge connection is a pure gauge.
 
\vspace{1.0cm}

{\noindent{\it{2.2: $n=4$ Conifold}}}

\vspace{0.5cm}

We now take up the conifold described by the quadric in $C^4$,
\begin{eqnarray}
\sum_{i=1}^{4} ({\phi}^i)^2 &=& 0.
\end{eqnarray}
This is a real-6 dimensional conifold. First, as in Section 2.1, we parameterize
the four complex fields (coordinates) satisfying (21) as,
\begin{eqnarray}
{\phi}^1\ =\ \psi(1+f_1f_2) &;& {\phi}^2\ =\ i\psi(1-f_1f_2), \nonumber \\
{\phi}^3 \ =\ \psi (f_1-f_2) &;& {\phi}^4 \ =\ i\psi(f_1+f_2.)
\end{eqnarray}
This parameterization is not exactly as in 
(3), but is more convenient and used in [21,22].  
For later use, we realize, as in [15], that 
by writing,
\begin{eqnarray}
Z&=& i{\phi}^4 I + \vec{\sigma} \cdot \vec{\phi}, \nonumber \\
 &=& 2 \left( \begin{array}{lccr} 
-f_2\psi & & & \psi \\
        &   & &  \\
(f_1f_2\psi) & & &  -f_1\psi \\
\end{array} \right), 
\end{eqnarray} 
we have an identification of the conifold with $SL(2,C)$.  
In (22), $\psi, f_1, f_2$ are three independent complex fields characterizing 
the conifold. The base of the conifold is given by the intersection of the 
space of solutions of (21), that is, (22), with a sphere in $C^4$,
\begin{eqnarray}
\sum_{i=1}^4 |{\phi}^i|^2 &=& 2 |\psi |^2 (1+ |f_1|^2)(1+ |f_2|^2)\ =\ r^2,
\ \ \ (say).
\end{eqnarray}
The apex of the cone is designated by $r=0$ which corresponds here to 
$\psi = 0$. The Ricci flat metric is constructed by following the strategy 
outlined before and it is given by,  
\begin{eqnarray}
g_{\alpha \bar{\beta}}&=&C\left( \begin{array}{l|c|r}
\frac{2ar^4}{3|\psi|^2} & \frac{4}{3}a\bar{\psi}f_1(1+|f_2|^2)r^2 &
\frac{4}{3} a\bar{\psi}f_2(1+|f_1|^2)r^2 \\
 & & \\
\frac{4}{3}a\psi {\bar{f}}_1(1+|f_2|^2)r^2 & \frac{2a|f_1|^2r^4/3+ar^4+b}
{(1+|f_1|^2)^2} & \frac{4}{3}a|\psi|^2{\bar{f}}_1f_2r^2 \\
 & & \\
\frac{4}{3}a \psi {\bar{f}}_2(1+|f_1|^2)r^2 & 
\frac{4}{3}a|\psi|^2f_1{\bar{f}}_2 r^2 
& \frac{2a|f_2|^2r^4/3+ar^4+b}{(1+|f_2|^2)^2} \\
\end{array} \right),
\end{eqnarray}
where $C\ =\ (ar^4+b)^{-\frac{2}{3}}$,  
$a$ and $b$ are integration constants and $\alpha, \beta$ stand for 
the fields $\psi, f_1, f_2$. The determinant can be verified as $\frac{8a}{3}
|\psi|^2$, so that the Ricci tensor $R_{\alpha \bar{\beta}}\ =\ -
{\partial}_{\alpha}{\partial}_{\bar{\beta}}\ell og det g$ is identically
zero.   

\vspace{0.5cm}

The metric (25) does not become singular   
as long as the integration constant $b \neq 0$}}. 
The apparent coordinate singularity when $\psi \rightarrow 0$ in the above
metric can be removed by redefining $\psi$ as $\Psi \ =\ \frac{1}{2} {
\psi}^2$. 
We  
note the appearance of the $b$ term in the $(22)$ and $(33)$ matrix elements 
of the metric (25). They correspond to 'adding sphere to the apex' of the 
conifold. 

\vspace{0.5cm}

The metric (25), upon using the parameterization, $\psi 
= \frac{r}{\sqrt{2}}e^{i\xi}cos{\frac{{\theta}_1}{2}}cos{\frac{{\theta}_2}
{2}}$, $f_1=e^{i{\phi}_1}tan{\frac{{\theta}_1}{2}}$ , $f_2=e^{i{\phi}_2}
tan{\frac{{\theta}_2}{2}}$, gives the line element of the 6-dimensional 
conifold as,
\begin{eqnarray}
(ds_6)^2&=&\frac{2}{3}ar^2\ (dr)^2 + \frac{1}{4}(ar^4+b) \Bigl\{ \sum_{i=1}
^{2}\Big( (d{\theta}_i)^2+sin^2{\theta}_i\ (d{\phi}_i)^2\Big)\Bigr\} 
\nonumber \\
&+&\frac{1}{6}ar^4{\Bigl\{ d\tilde{\xi} + cos{\theta}_1\ d{\phi}_1 +
 cos{\theta}_2\ d{\phi}_2 \Bigr\} }^2,   
\end{eqnarray}
where we have introduced $\tilde{\xi}=-2\xi -{\phi}_1-{\phi}_2$.    

\vspace{0.5cm}

A non-linear sigma model on the target
space as the conifold (25) is described by the action,
\begin{eqnarray}
S&=& \int g_{\alpha \bar{\beta}} \{ {\partial}_z{\alpha} {\partial}_{\bar{z}}
{\bar{\beta}}+{\partial}_{\bar{z}}{\alpha}{\partial}_{z}{\bar{\beta}}\} 
\frac{i}{2} dz d\bar{z},
\end{eqnarray}
where $\alpha, \beta$ stand for $\psi, f_1, f_2$ and the summation over them
is understood. The closed K\"{a}hler 2-form $\omega$ is used to write the 
topological invariant,
\begin{eqnarray}
\int \omega &=& -2i \int g_{\alpha \bar{\beta}} \{ {\partial}_z \alpha 
{\partial}_{\bar{z}}\bar{\beta} - {\partial}_{\bar{z}}\alpha {\partial}_z
\bar{\beta}\} \frac{i}{2} dz d\bar{z},
\end{eqnarray}
so that we have,
\begin{eqnarray}
S+c\int \omega &=& 2\int g_{\alpha \bar{\beta}}{\partial}_z\alpha {\partial}
_{\bar{z}}\bar{\beta} \frac{i}{2} dz d\bar{z}.
\end{eqnarray}

\vspace{0.5cm}

Using the metric components in (25), for $b=0$, we find,
\begin{eqnarray}
S+c\int \omega& =& 2\int \Bigl\{ \left | {\psi}_z+\frac{{\bar{f}}_1f_{1z}}{(1+
|f_1|^2)}\psi + \frac{{\bar{f}}_2f_{2z}}{(1+|f_2|^2)}\psi\right |^2 \nonumber \\
&+&(ar^4+b)\Bigl\{ \frac{|f_{1z}|^2}{(1+|f_1|^2)^2}+\frac{|f_{2z}|^2}{
(1+|f_2|^2)^2}\Bigr\} \Bigr\}\frac{i}{2}\ dz d\bar{z}, \nonumber \\
&\geq & 0.
\end{eqnarray}
Similar procedure for $S-c\int \omega$ can be combined to get the result that 
the sigma model action on the conifold satisfies the inequality,
\begin{eqnarray}
S&\geq |c\int \omega |.
\end{eqnarray}
The equality holds good when the fields $\psi, f_1, f_2$ are either 
holomorphic  or anti-holomorphic, 
 in which case the minimum of the action will be 
a topological invariant.  
 
\vspace{0.5cm}

The holomorphic classical background fields  
which are solutions to the equations of motion for the minimum action in
(31), 
can be 
generically taken as $P(z)/Q(z)\ = \ \frac{pz+q}{sz+t}$, as 1-instanton 
configuration. Then there is a structural similarity between the conifold 
(21,22,23) and the moduli space of 1-instanton as,
\begin{eqnarray}
Z\ =\ \left( \begin{array}{lcr}
-2f_2\psi & & 2\psi \\
 & & \\
(2f_1f_2\psi) & & -2f_1 \psi \\
\end{array} \right) 
&\leftrightarrow & \left( \begin{array}{lcr}
p & & q \\
 & & \\
s & & t \\
\end{array} \right). 
\end{eqnarray}
This is recently pointed out by Speight [12].

\vspace{0.5cm}

The limit $\psi \rightarrow 0$ of the minimum action in (31) corresponds to
Grassmannian non-linear sigma model which was extensively studied in [21] 
as describing QCD strings. 

\vspace{0.5cm}

{\noindent{\bf{3. Resolved Conifold}}} 

\vspace{0.5cm}

We now consider the case of a resolved conifold. 
 The metric for the resolved conifold, in terms of 
real coordinates, has been obtained by Pando Zayas and Tseytlin [27]. We will 
describe it here {\it{in terms of complex coordinates}} which will be 
convenient for our purpose of describing sigma model. 
Nevertheless, we will 
give, for comparison with [27], the metric in terms of real coordinates. 
We follow [15] for the construction of a resolved conifold.

\vspace{0.5cm}

We consider (21, 22, 23) of Section 2.2 for $n=4$ conifold and realize that 
(21) is equivalent to $det Z\ =\ 0$. Defining $W\ =\ \frac{1}{\sqrt{2}}Z$, and
dispensing $\psi$ (which will be reintroduced shortly as $\lambda$), 
we have,
\begin{eqnarray}
W&=& \sqrt{2} \left( \begin{array}{lcr}
 -f_2 & & 1 \\
 & &  \\
(f_1f_2) & & -f_1 \\
\end{array} \right) \ \equiv \ \left( \begin{array}{lcr}
X & & U \\
 & & \\
V & & Y \\
\end{array} \right), 
\end{eqnarray} 
where $X=-\sqrt{2}f_2, U=\sqrt{2}, V=\sqrt{2}f_1f_2, Y=-\sqrt{2}f_1$ and 
$XY-UV\ =\ 0$. The resolved conifold is obtained [15] by replacing the
preceeding relation by a pair of equations,
\begin{eqnarray}
\left( \begin{array}{lcr}
X & & U \\
 & & \\
V & & Y \\
\end{array} \right) \ \left( \begin{array}{c}
{\lambda}_1 \\
  \\
{\lambda}_2 \\
\end{array} \right) &=& 0,
\end{eqnarray}
where $({\lambda}_1, {\lambda}_2)$ are not both zero. Equations,(34)
describe $C_4\times P_1$, with the node having been replaced by a $P_1
=S^2$. We  
will work in the region ${\lambda}_2 \neq 0$ and define $\lambda \ =\ 
\frac{{\lambda}_1}{{\lambda}_2}$. A solution to the above equation is given 
by $U=-X\lambda, \ Y=-V\lambda$. Then, it follows that,
\begin{eqnarray}
W&=& \sqrt{2}f_2 \left( \begin{array}{lcr}
-1 & & \lambda \\
  & & \\
f_1 & & (-f_1\lambda) \\
\end{array} \right). 
\end{eqnarray} 
Thus $f_1,f_2,\lambda$ are the three complex coordinates characterizing the
resolved conifold {\it{in the patch ${\lambda}_2 \neq 0$}}. The $O(4)$  
invariant quantity is given (see Section:2.2 for comparison) by 
\begin{eqnarray}
r^2 &=& Tr(W^{\dagger}W)\ =\ 2| f_2|^2 (1+|f_1|^2)(1+| \lambda |^2),
\end{eqnarray}
which is very similar to (24).  
In here, the scalar differentiable 
function $K$ 
is taken to be,
\begin{eqnarray}
K &=& F(r^2) + 4a^2 \ell og (1+| \lambda |^2).
\end{eqnarray}
That is, in addition to the $\lambda $ dependence through $r^2$,  
we have an additional term that
depends on $\lambda$ only.  
In the limit $a\rightarrow 0$, we obtain the 
ordinary conifold. The function $F$ is determined by the same strategy used 
earlier. 

\vspace{0.5cm}

Working out the metric as $g_{\alpha \bar{\beta}}\ =\ {\partial}_{\alpha}
{\partial}_{\bar{\beta}}K$, we find,
\begin{eqnarray}
g_{\lambda \bar{\lambda}}&=&2|f_2|^2(1+|f_1|^2)\{ 2F''|\lambda|^2|f_2|^2
(1+|f_1|^2)+F'\}+\frac{4a^2}{(1+|\lambda|^2)^2}, \nonumber \\
g_{f_1\bar{\lambda}}&=&2\lambda {\bar{f}}_1|f_2|^2(r^2F''+F'), \nonumber \\
g_{f_2\bar{\lambda}}&=&2\lambda {\bar{f}}_2(1+|f_1|^2)(r^2F''+F'), 
\nonumber \\
g_{f_1{\bar{f}}_1}&=&2|f_2|^2(1+|\lambda |^2)\{2|f_1|^2|f_2|^2(1+|\lambda
 |^2)F''+F'\}, \nonumber \\
g_{f_1{\bar{f}}_2}&=&2{\bar{f}}_1f_2(1+|\lambda |^2)(r^2F''+F'), \nonumber \\
g_{f_2{\bar{f}}_2}&=&2(1+|\lambda |^2)(1+|f_1|^2)(r^2F''+F'),
\end{eqnarray}
where $F', F''$ represent first and second order differential of $F$ with 
respect to $r^2$. The determinant of this metric is found to be,
\begin{eqnarray}
detg &=& 4|f_2|^2F'(r^2F''+F')(r^2F'+4a^2),
\end{eqnarray}
agreeing with [27]. In order to realize Ricci flatness, we need to just equate
from (39), 
\begin{eqnarray}
F'(r^2F''+F')(r^2F'+4a^2) &=& constant.
\end{eqnarray}
The same equation is obtained in [27] and this ensures that our parameterization 
is correct. By letting $r^2F'\ =\ \gamma$ and after one integration, we convert
the above non-linear equation to a cubic algebraic equation,
\begin{eqnarray}
{\gamma}^3+6a^2{\gamma}^2-r^4 &=& 0,
\end{eqnarray}
where the constant in (39) is taken as $2/3$. The real solution of this is 
given by,
\begin{eqnarray}
\gamma &=& s_1+s_2-2a^2, \nonumber \\
s_{1,2} &=& {\Bigl\{\frac{r^4}{2}-8a^6 \pm \frac{1}{2} \sqrt{{r^8}-
32a^6r^4}\Bigr\}}^
{\frac{1}{3}}.
\end{eqnarray}
Realizing that $s_1s_2\ =\ 4a^4$, the solution can be written as,
\begin{eqnarray}
\gamma &=& -2a^2+4a^2{s_1}^{-1}+s_1.
\end{eqnarray}
and so,
\begin{eqnarray}
F' &=& \frac{\gamma}{r^2}, \nonumber \\
F'' &=& \frac{2}{3\gamma (\gamma + 4a^2)}-\frac{\gamma}{r^4}.
\end{eqnarray}
Thus, the components of the Ricci flat metric in the complex basis for the 
resolved conifold are,
\begin{eqnarray}
g_{\lambda {\bar{\lambda}}}&=&\frac{1}{(1+| \lambda |^2)^2}\Bigl\{
\frac{2r^4}{3\gamma(\gamma + 4a^2)}| \lambda |^2+\gamma + 4a^2 \Bigr\}, 
\nonumber \\
g_{f_1\bar{\lambda}}&=&\frac{\lambda {\bar{f}}_1}{(1+| f_1 |^2)(1+
| \lambda |^2)}\ \frac{2r^4}{3\gamma (\gamma + 4a^2)}, \nonumber \\
g_{f_2 \bar{\lambda}}&=&\frac{\lambda {\bar{f}}_2}{| f_2 |^2(1+| \lambda |^2)}
\ \frac{2r^4}{3\gamma (\gamma + 4a^2)}, \nonumber \\
g_{f_1 {\bar{f}}_1}&=&\frac{1}{(1+|f_1|^2)^2}\{ \frac{2r^4}{3\gamma(\gamma 
+ 4a^2)}|f_1 |^2 + \gamma \}, \nonumber \\
g_{f_1{\bar{f}}_2}&=&\frac{{\bar{f}}_1f_2}{(1+|f_1 |^2)| f_2 |^2}\ 
\frac{2r^4}{3\gamma(\gamma + 4a^2)}, \nonumber \\
g_{f_2 {\bar{f}}_2} &=& \frac{2r^4}{3\gamma (\gamma + 4a^2)|f_2 |^2}.
\end{eqnarray}  
 
\vspace{0.5cm}

Before proceeding to a non-linear sigma model on this resolved conifold, we 
give a realization of the above metric in terms of real coordinates.  
Parameterizing the complex coordinates $f_1,f_2$ and $\lambda$, in a 
manner slightly from the one that led to (26), as,
\begin{eqnarray}
f_1 &=& e^{-i{\phi}_1} tan{\frac{{\theta}_1}{2}}, \nonumber \\
f_2 &=& \frac{r}{\sqrt{2}} e^{\frac{i}{2}\psi} cos{\frac{{\theta}_1}{2}}
cos{\frac{{\theta}_2}{2}}, \nonumber \\
\lambda &=& e^{-i{\phi}_2} tan {\frac{{\theta}_2}{2}},
\end{eqnarray}
such that $2|f_2|^2(1+|f_1|^2)(1+| \lambda |^2)\ =\ r^2$,  
we find, 
\begin{eqnarray}
(ds_6)^2 &=& g_{\alpha \bar{\beta}} d\alpha d\bar{\beta}, \nonumber \\
 &=& {\gamma}'(dr)^2+r^2{\gamma}'\frac{1}{4} {\Bigl\{ d\tilde{\psi} 
- \sum_{i=1}
^{2} cos{\theta}_i\ d{\phi}_i \Bigr\} }^2 \nonumber \\
&+& \frac{\gamma}{4} \sum_{i=1}^{2} \Bigl\{ (d{\theta}_i)^2 + sin^2
{\theta}_i \ (d{\phi}_i)^2\Bigr\} \nonumber \\
&+& a^2 \{ (d{\theta}_2)^2 + sin^2{\theta}_2\ (d{\phi}_2)^2\},
\end{eqnarray} 
the metric on the resolved conifold, agreeing with [27] after the 
redefinition $\tilde{\psi}=\psi -{\phi}_1-{\phi}_2$. 
For completeness, we rewrite this 
metric first by introducing ${\rho}^2 \ =\ \frac{3}{2}\gamma $ and then 
treat $\rho$ as the radial coordinate, instead of $r$. This leads to the 
metric as,
\begin{eqnarray}
(ds_6)^2&=& {\kappa (\rho)}^{-1} (d\rho )^2 + \frac{1}{9} \kappa (\rho)
{\rho}^2 {e_{\tilde{\psi}}}^2 + \frac{{\rho}^2}{6}( {e_{{\theta}_1}}^2+
{e_{{\phi}_1}}^2) \nonumber \\
&+& (\frac{{\rho}^2}{6}+a^2)({e_{{\theta}_2}}^2+{e_{{\phi}_2}}^2),
\end{eqnarray}
where $e_{\psi}\ =\ \{ d\tilde{\psi} + \sum_{i=1}^{2}
cos{\theta}_i\ d{\phi}_i\}$, $e_{{\theta}_i}\ =\ d{\theta}_i$,  
$e_{{\phi}_i}\ =\ sin {\theta}_i\ d{\phi}_i$ and $\kappa (\rho)=
\frac{{\rho}^2+9a^2}{{\rho}^2+6a^2}$.    

\vspace{0.5cm}

We point out two uses of this metric and its complex realization (45) in the 
context of D-branes. First, given the Ricci flat metric on the transverse 
six dimensional space taken as a conifold, the standard brane solution 
[3,27] is
\begin{eqnarray}
(ds_{10})^2&=& H^{-\frac{1}{2}}(y) {\eta}_{\mu\nu}dx^{\mu}dx^{\nu} + 
H^{\frac{1}{2}}(y) g_{ij}dy^idy^j,
\end{eqnarray}
as a warped metric, where $y$ collectively denotes the coordinates of the 
transverse six dimensional space and  
the warp factor $H(y)$ is a harmonic  in
the transverse space, that is,
\begin{eqnarray}
\frac{1}{\sqrt{g}}{\partial}_i(\sqrt{g} g^{ij}{\partial}_jH)&=&0.
\end{eqnarray}
Using the metric given in (48) and assuming that  
$H$  
 depends on $\rho$ only, one can solve (50) for $H(\rho)$ as,  
\begin{eqnarray}
H(\rho) &=& H_0+C\Bigl\{ \frac{1}{18a^2{\rho}^2} - \frac{1}{162a^4} \ell og (
1+ \frac{9a^2}{{\rho}^2})\Bigr\}.
\end{eqnarray}

For small values of $\rho$, it follows that $H(\rho) \rightarrow \frac{
C}{18a^2{\rho}^2}$ and $\kappa (\rho) \rightarrow \frac{3}{2}$, so that 
the brane solution becomes,
\begin{eqnarray}
(ds_{10})^2 &=& \frac{3\sqrt{2}a\rho}{\sqrt{C}}{\eta}_{\mu\nu}dx^{\mu}
dx^{\nu} + \frac{2\sqrt{C}}{9\sqrt{2}a} \ \frac{1}{\rho} (d\rho )^2 
\nonumber \\
&+& \frac{\sqrt{C}}{3\sqrt{2}a}\Bigl\{ \frac{1}{6}\rho (e^2_{\psi} +
e^2_{{\theta}_1}+e^2_{{\theta}_2}+e^2_{{\phi}_1}+e^2_{{\phi}_2}) 
+\frac{a^2}{9}(e^2_{{\theta}_2}+e^2_{{\phi}_2})\Bigr\}. 
\end{eqnarray}
This solution corresponds to $AdS_5 \times X_5$.  

Second, we have seen that the complex coordinates ${\phi}^i$ in (21) 
transform in the four dimensional representation of $SO(4)$ and have 
unit charge with respect the $U(1)$ symmetry group. Klebanov and Witten [3]
have given a holomorphic three form as,
\begin{eqnarray}
\Omega &=& \frac{d{\phi}^2\wedge d{\phi}^3\wedge d{\phi}^4}{{\phi}^1},
\end{eqnarray}
which in our complex parameterization (22) becomes $\Omega \ =\ -2\psi \ 
d\psi \wedge df_1 \wedge df_2$ and has "charge two" under the said $U(1)$. 
This will be an $R$-symmetry group. 

\vspace{0.5cm}

Now we will consider a 2-dimensional non-linear sigma model with the target 
space as the above (complex version) resloved conifold. Denoting
$\frac{2r^4}{3\gamma(\gamma + 4a^2)}$ by $\Gamma (r^2)$, the action becomes,
\begin{eqnarray}
S&=& \int \Bigl\{ \Big( \frac{1}{(1+| \lambda |^2)^2}(| \lambda |^2\Gamma 
+\gamma) + \frac{4a^2}{(1+| \lambda |^2)^2}\Bigr) ({\lambda}_z
{\bar{\lambda}}_{\bar{z}}+{\lambda}_{\bar{z}}{\bar{\lambda}}_z) 
\nonumber \\
&+&\frac{\lambda {\bar{f}}_1\Gamma}{(1+|f_1|^2)(1+|\lambda |^2)}(f_{1z}
{\bar{\lambda}}_{\bar{z}}+f_{1\bar{z}}{\bar{\lambda}}_z) + h.c 
\nonumber \\
&+& \frac{\lambda {\bar{f}}_2\Gamma}{|f_2|^2(1+|\lambda |^2)}(f_{2z}
{\bar{\lambda}}_{\bar{z}}+{f}_{2\bar{z}}{\bar{\lambda}}_z) + h.c 
\nonumber \\
&+&\frac{1}{(1+|f_1|^2)^2}(\Gamma |f_1|^2 + \gamma)(f_{1z}{\bar{f}}_{1
\bar{z}}+f_{1\bar{z}}{\bar{f}}_{1z}) \nonumber \\
&+&\frac{{\bar{f}}_1f_2\Gamma}{|f_2|^2(1+|f_1|^2)}(f_{1z}{\bar{f}}_
{2\bar{z}}+f_{1\bar{z}}{\bar{f}}_{2z}) + h.c 
\nonumber \\
&+& \frac{\Gamma}{|f_2|^2}(f_{2z}{\bar{f}}_{2\bar{z}}+f_{2\bar{z}}
{\bar{f}}_{2z}) \Bigr\} \Bigl( \frac{i}{2} \Bigr) dz d\bar{z}.
\end{eqnarray}

\vspace{0.5cm}

The K\"{a}hler 2-form $\omega$ can be similarly written and is found to be 
closed. In order to investigate whether the above action has a minimum, we
consider the sum (and difference as well) which turns out to be,
\begin{eqnarray}
S+c\int \omega &=& 2\int \Bigl\{ \frac{(4a^2+\gamma)}{(1+|\lambda |^2)^2}
|{\lambda}_z|^2 + \frac{\gamma}{(1+|f_1|^2)^2} |f_{1z}|^2 \nonumber \\
&+& \frac{\Gamma |\lambda |^2}{(1+|\lambda |^2)^2} \left | {\lambda}_z + 
\frac{(1+|\lambda|^2)\lambda {\bar{f}}_1}{|\lambda|^2(1+|f_1|^2)}f_{1z}
+\frac{\lambda (1+|\lambda |^2){\bar{f}}_2}{|\lambda |^2|f_2|^2}f_{2z}
\right |^2 \Bigr\} \Big( \frac{i}{2}\Big) dz d\bar{z} \nonumber \\
&\geq & 0.
\end{eqnarray}
Thus, the non-linear sigma model action with the target space as the resolved 
conifold is bounded below.  
This action has a smooth behaviour and the equations of motion of the 
minimum action do not become singular. The structural 
similarity between the 
resolved conifold and the moduli space of unit charge instanton (which are 
the solutions of the equations of motion) is easily seen as in the case of the
ordinary conifold.   

\vspace{0.5cm}

In contrast to the situation in the non-linear sigma model with the ordinary 
conifold as the target space, here, the role of the gauge connection needs 
 to be analysed only for the second line in (55). The first 
term gives the action on $S^2$ which essentially replaces the apex. Considering
the second line, the covariant derivative can be easily noticed with the 
gauge connection identified with the $U(1)$ fibre on the base of the resolved
conifold. As the target space is Ricci flat, perturbative 1-loop corrections
are absent and the action becomes topological.  

\vspace{0.5cm}

{\noindent{\bf{4. Conclusion}}}

\vspace{0.5cm}

We have constructed Ricci-flat K\"{a}hler metric for the conifold of complex 
dimensions $n=3,4$ in terms of complex 
coordinates.  
This complements the study using real coordinates. The strategy 
followed consisted in solving the differential equation for the K\"{a}hler 
potential. With the choice of integration constant $b$ not set zero, the 
metric remains smooth. A realization of the metric 
in terms of real coordinates is made for both $n=3,4$ conifolds. 
With the intention 
of constructing a field theory on conifold, we considered 
two dimensional non-linear sigma 
model on the conifold, by identifying the  complex coordinates as 
sigma model fields defined on a 2-dimensional space.  
The closed K\"{a}hler 2-form is used to obtain a lower bound for the 
action for the non-linear sigma model. The minimum action 
corresponds to the complex fields being either holomorphic or 
anti-holomorphic in the complex 2-dimensional space. The classical equations 
of motion are found to be non-singular, by the choice of the integration 
constant. This suggests a method to overcome the difficulties in deriving 
a low energy effective action in the case of Calabi-Yau compactification. 

\vspace{0.5cm}

The same procedure of using complex coordinates is extended to find the 
metric (in the complex basis) for the $n=4$ resolved conifold 
and its realization  
in terms of six real coordinates is made. This agrees with [27]. The 
harmonic function appearing as the warp factor in the solution of Type-IIB 
string theory is determined and in the $\rho \rightarrow 0$ limit,  
this solution goes over to  $AdS_5 
\times X_5$ geometry. A non-linear sigma model on the resolved conifold 
is constructed using our complex realization of the resolved conifold 
metric.  

\vspace{0.5cm}

{\noindent{\bf{Acknowledgements}}

\vspace{0.5cm}

One of the authors (R.P) wishes to acknowledge with thanks the hospitality 
at the Simon Fraser University. Useful discussions with R.C.Rashkov and 
P.Matlock are thankfully acknowledged. This work is supported in part by an 
operating grant from the Natural Sciences and Engineering Research Council 
of Canada.

\vspace{0.5cm}

{\noindent{\bf{References}}}

\vspace{0.5cm}

\begin{enumerate}

\item J.Maldacena, {\it{The large N limit of Superconformal field theories 
                   and Supergravity}}, Adv.Theor.Math.Phys. {\bf{2}} (1998)
                   231; hep-th/9711200.  
\item E.Witten, {\it{Anti-de Sitter space and holography}}, hep-th/9802150.  
\item I.R.Klebanov and E.Witten, {\it{Superconformal Field Theory on Three 
                   Branes at a Calabi-Yau Singularity}}, hep-th/9807080. 
\item L.Romans, Phys.Lett. {\bf{B153}} (1985) 392. 
\item D.Page and C.Pope, Phys.Lett. {\bf{B144}} (1984) 346.  
\item I.R,Klebanov and M.J.Strassler, {\it{Supergravity and a confining 
                  gauge theory: duality cascades and $\chi$SB-resolution of 
                  naked singularity}}, JHEP {\bf{0008}} (2000) 052. 
\item C.P.Herzog, I.R.Klebanov and P.Ouyang, {\it{Remarks on the warped 
                  deformed conifold}}- hep-th/0108101.
\item R.Minasian and D.Tsimpis, {\it{On the geometry of non-trivially 
                  embedded branes}}, hep-th/9911092.
\item R.Unge, {\it{Branes at generalized conifolds and toric geometry}}, 
               hep-th/9901091.
\item G.Papadopoulos and A.A.Tseytlin, {\it{Complex geometry of conifolds 
                 and 5-Branes wrapped on a 2-sphere}}, hep-th/0012034.
\item S.Gukov, C.Vafa and E.Witten, {\it{CFT from Calabi-Yau 4-folds}}, 
               hep-th/9906070.
\item J.M.Speight, {\it{The deformed conifolds as a geometry on the space 
                of unit charge $CP^1$-lumps}}, hep-th/0105142.
\item E.Caceres and R.Hernandez, {\it{Wilson loops in the Higgs phase of 
               Large N Field theories on the conifold}}, hep-th/0004040.
\item A.Strominger, {\it{Massless blackholes and conifolds in string theory}},
               hep-th/9504090.
\item P.Candelas and X.C.de la Ossa, Nucl.Phys. {\bf{B342}} (1990) 246.
\item M.B.Stenzel, Manuscripta Mathematica, {\bf{80}} (1993) 151.
\item G.W.Gibbons and C.N.Pope, Comm.Math.Phys. {\bf{66}} (1979) 267.
\item M.Cvetic, G.W.Gibbons, H.Lu and C.N.Pope, {\it{Ricci-flat metrics, 
              Harmonic forms and Brane resolutions}}, hep-th/0012011.
\item K.Higashijima, T.Kimura and M.Nitta, Phys.Lett. {\bf{B518}} (2001) 301. 
\item A.Kaya, {\it{On conifolds and D3-branes}}, hep-th/0110214.
\item K.S.Viswanathan, R.Parthasarathy and D.Kay, Ann.Phys.(NY) {\bf{206}} 
              (1991) 237, \ 
      K.S.Viswanathan and R.Parthasarathy, Phys.Rev. {\bf{D51}} (1995) 5830; 
             Ann.Phys. (NY) {\bf{244}} (1995) 241; Phys.Rev. {\bf{D55}} 
             (1997) 3800. \
      R.Parthasarathy and K.S.Viswanathan, Int.J.Mod.Phys. {\bf{A7}} (1992) 
             1819.
\item D.A.Hoffman and R.Osserman, J.Diff.Geom. {\bf{18}} (1983) 733; Proc. 
         London. Math.Soc. (3) {\bf{50}} (1985) 21.
\item T.J.Willmore, {\it{Riemannian Geometry}}, Clarendon Press, Oxford, 
        1993.
\item A.M.Perelomov, Comm.Math.Phys. {\bf{63}} (1978) 237.
\item V.A.Fateev, I.V.Frolov and A.S.Schwarz, Nucl.Phys. {\bf{B154}} 
       (1979) 1.
\item L.Alvarez-Gaume, D.Z.Freedman, and S.Mukhi, Ann.Phys.(NY) {\bf{134}} 
       (1981) 85.
\item L.A.Pando Zayas and A.A.Tseytlin, {\it{3-Branes on Resolved Conifold}},
       hep-th/0010088.
\end{enumerate}     

\end{document}